# High-resolution neural network-driven mapping of multiple diffusion metrics leveraging asymmetries in the balanced SSFP frequency profile

*Florian Birk[1], Felix Glang[1], Alexander Loktyushin[1,2], Christoph Birkl[3], Philipp Ehses[4], Klaus Scheffler[1,5], and Rahel Heule[1,5]*

[1]High Field Magnetic Resonance, Max Planck Institute for Biological Cybernetics, Tübingen, Germany.

[2]Empirical Inference, Max Planck Institute for Intelligent Systems, Tübingen, Germany.

[3]Department of Neuroradiology, Medical University of Innsbruck, Innsbruck, Austria.

[4]German Center for Neurodegenerative Diseases (DZNE), Bonn, Germany.

[5]Department of Biomedical Magnetic Resonance, University of Tübingen, Tübingen, Germany.

**Correspondence to:**

Rahel Heule, PhD

High Field Magnetic Resonance

Max Planck Institute for Biological Cybernetics

Max-Planck-Ring 11

72076 Tübingen, Germany

Email: rahel.heule@tuebingen.mpg.de

Phone: +49-7071-601-704

Fax:    +49-7071-601-702

**Funding information**

Max Planck Society; German Research Foundation (DFG, Reinhart Koselleck Project, DFG SCHE 658/12)

# ABSTRACT

We suggest to utilize the rich information content about microstructural tissue properties entangled in asymmetric balanced steady-state free precession (bSSFP) profiles to estimate multiple diffusion metrics simultaneously by neural network (NN) parameter quantification. A 12-point bSSFP phase-cycling scheme with high-resolution whole-brain coverage is employed at 3 T and 9.4 T for NN input. Low-resolution target diffusion data are derived based on diffusion-weighted spin-echo echo-planar-imaging (SE-EPI) scans, i.e., mean, axial, and radial diffusivity (MD, AD, RD), fractional anisotropy (FA) as well as the spherical coordinates (azimuth $\Phi$ and inclination $\Theta$) of the principal diffusion eigenvector. A feedforward NN is trained with incorporated probabilistic uncertainty estimation.

The NN predictions yielded highly reliable results in white matter (WM) and gray matter (GM) structures for MD. The quantification of FA, AD, and RD was overall in good agreement with the reference but the dependence of these parameters on WM anisotropy was somewhat biased, e.g., in corpus callosum. The inclination $\Theta$ was well predicted for anisotropic WM structures while the azimuth $\Phi$ was overall poorly predicted. The findings were highly consistent across both field strengths. Application of the optimized NN to high-resolution input data provided whole-brain maps with rich structural details. In conclusion, the proposed NN-driven approach showed potential to provide distortion-free high-resolution whole-brain maps of multiple diffusion metrics at high to ultra-high field strengths in clinically relevant scan times.

**Abbreviations used:** AD, axial diffusivity; AI, asymmetry index; $B_0$, static magnetic field; bSSFP, balanced steady-state free precession; CC, corpus callosum; CNN, convolutional neural network; CSF, cerebrospinal fluid; DTI, diffusion tensor imaging; DWI, diffusion-weighted imaging; EPI, echo-planar imaging; FA, fractional anisotropy; GM, gray matter; GRAPPA, generalized autocalibrating partial parallel acquisition; HL, hidden layer; IC, internal capsule; MD, mean diffusivity; MPRAGE, magnetization-prepared rapid gradient echo; MSE, mean squared error; NN, neural network; OR, optic radiation; PSF, point spread function; RD, radial diffusivity; ReLU, rectified linear unit; ROI, region-of-interest; SE, spin echo; SD, standard deviation; SNR, signal-to-noise ratio; SSFP, steady-state free precession; UHF, ultra-high field; WM, white matter.

# INTRODUCTION

Diffusion-weighted imaging (DWI) allows measuring the directionality of microstructural diffusion processes and delineating white matter (WM) fiber pathways based on the anisotropic movement of water protons. The acquired MR signal is sensitized to the diffusion of water molecules in biological tissues by the application of strong magnetic field gradients in multiple directions. High-resolution DWI holds promise to resolve brain tissue structures at a microscopic level, aiding in the visualization of fine-scale axonal fiber architecture in WM[1] or the detection of cortical gray matter (GM)[2] anisotropy. Considering the complexity of diffusion processes underlying brain tissue microstructure, DWI is expected to greatly benefit from imaging at high isotropic submillimeter resolution.

DWI is a widely used MR protocol in clinical disease assessment and applied to reveal brain tissue changes following acute ischemic strokes[3–5], but also to trace degenerative diseases like multiple sclerosis[6–8], Parkinson's disease[9,10] or Alzheimer's disease[11–13]. Quantitative metrics derived from diffusion tensor imaging (DTI) such as mean diffusivity (MD) or fractional anisotropy (FA) reflect alterations in the tissue microenvironment. Increased MD values are an indicator for damaged tissues since lower cell integrity leads to increased free diffusion[3]. Reduced FA, on the other hand, is a well-known marker of axon loss and declines in myelin integrity[3,7,13], especially for WM. Axial diffusivity (AD) and radial diffusivity (RD) measure water diffusion parallel and perpendicular to axon bundles, providing increased specificity to differentiate between axonal damage and myelin injury (demyelination), respectively, as reported within studies in mice[14–17] and humans[18–20]. This interpretation based on the underlying biophysical properties such as axon or myelin status may, however, fail in regions of complex brain tissue architecture, e.g., in voxels characterized by crossing fibers[21].

Most current DWI sequences are based on the original Stejskal-Tanner pulsed-gradient spin-echo (SE) technique[22]. Strong monopolar or bipolar diffusion-sensitizing gradients are applied during the SE module, followed by a fast single- or multi-shot echo-planar-imaging (EPI) readout. Diffusion-weighted SE-EPI is yet prevailing, but it has a number of limitations related to its typically low bandwidth in phase encoding direction, in particular severe image distortions in the presence of static magnetic field ($B_0$) inhomogeneity and spatial blurring due to $T_2^*$ decay along the echo train. These image degradations can be mitigated to a certain degree at the cost of prolonged scan times in multi-shot segmented EPI acquisitions provided that dedicated (self-)navigation techniques are applied to remove motion-induced shot-to-shot phase instabilities[23–27].

The issue of geometric distortion and $T_2^*$-blurring in echo-planar DWI is amplified at ultra-high fields due to larger $B_0$ field variations causing a loss of the intrinsic spatial resolution. In addition, the relative level of distortion with respect to the voxel size increases at higher resolutions. Recently, a diffusion-weighted point spread function (PSF) mapping approach with multi-shot SE-EPI encoding was suggested to achieve high-resolution distortion-free diffusion data at 7 T[28]. However, isotropic high-resolution diffusion imaging providing volumetric brain coverage within clinically relevant measurement times remains challenging and hardly feasible at ultra-high field strength. Steady-state free precession (SSFP) imaging is a promising alternative, which is not prone to geometric distortions and suited for 3D scans at high isotropic submillimeter resolution. Efficient diffusion weighting can be achieved by adding diffusion-encoding gradients to each cycle in the rapid RF pulse train[29]. However, SSFP with unbalanced gradient waveforms is vulnerable to motion artifacts, hindering its application to *in vivo* brain imaging, but its potential has been explored for *ex vivo* DTI in fixed human brain tissue[30,31].

Interestingly, it was reported that information about tissue anisotropies is entangled in the phase-cycled balanced SSFP (bSSFP) signal, manifest by frequency responses with different degrees of asymmetry depending on the underlying tissue type and the corresponding intravoxel frequency distribution[32–34]. Miller et al. observed pronounced bSSFP profile asymmetries in WM and compared the asymmetry level with DTI data[33]. The respective findings suggest significant correlations of the bSSFP asymmetry index (AI) with DTI metrics, such as the component of the principal diffusion eigenvector parallel to $B_0$, providing information about fiber tract orientation, or FA, reflecting the strength of tract directionality. Highest bSSFP asymmetries are found in highly anisotropic WM tracts oriented perpendicular to $B_0$[33]. The results of a recent study investigating the contribution of chemical exchange effects to asymmetric intravoxel frequency distributions in WM corroborate that the main factor driving bSSFP profile asymmetries in WM is structure- and not exchange-related[35].

This inherent sensitivity to tissue microstructure, combined with a mixed dependence on both $T_1$ and $T_2$, as well as the ability to enable distortion-free motion-robust volumetric imaging with high SNR efficiency in short scan times[36,37] make phase-cycled bSSFP an interesting tool for multi-parametric mapping of various MR quantities[38–40]. Since off-resonance effects in the bSSFP signal arising from $B_0$ inhomogeneities are used as an additional encoding dimension for multi-parametric mapping, phase-cycled bSSFP is well suited for high-resolution imaging at ultra-high field strength. However, current quantitative methods using phase-cycled bSSFP data, do not include the intravoxel frequency dispersion in the employed analytical signal model. This results, e.g., in a considerable bias of $T_1$ and $T_2$ quantification in brain tissues, especially in WM[38,39], which can be eliminated by a deep learning-driven approach as reported recently[40].

Here, we propose to utilize the rich information content about tissue microstructure entangled in bSSFP profile asymmetries to simultaneously estimate multiple diffusion measures by means of artificial neural networks (NNs). Concretely, we aim at quantifying the scalar metrics MD, FA, AD, RD as well as the spherical coordinates Φ (azimuth) and Θ (inclination) of the principal diffusion eigenvector directly from 12-point 3D phase-cycled bSSFP input data acquired at high (3 T) and ultra-high (9.4 T) field strength in healthy volunteers. To this end, a feedforward NN is investigated at each field strength with incorporated probabilistic uncertainty estimation offering the ability to quantify the confidence level of the predicted parameters in each voxel[41–43]. Target data are obtained at 3 T using a standard 2D multi-slice SE-EPI DTI measurement.

**METHODS**

MRI experiments were performed on healthy subjects at a field strength of 3 T (Magnetom Prisma, Siemens Healthineers, Erlangen, Germany) using a manufacturer-built 64-channel receive head array coil and at 9.4 T (Siemens Magnetom) using a custom-built head coil array consisting of 18 transceiver surface loops and 14 receive-only vertical loops[44]. All measurements conducted in this study were in accordance with the local ethical guidelines and each subject gave written informed consent before scanning. Image registration and segmentation as well as DTI fitting including distortion correction were performed with the software packages *FSL*[45] and *AFNI*[46]. Data preparation for the NN and visualization of the results were implemented using *Matlab* (R2019b, The MathWorks, Inc., Natick, MA). The open-source *Python*-based deep-learning framework *Keras*[47] (version 2.3.1), with the *TensorFlow*[48] (version 2.2.0) backend, was used to train the NN models. The *Python* package *scikit-optimize*[49] enabled hyperparameter optimization to find adequate NN architectures.

Data acquisition at 3 T

Whole-brain 3D sagittal bSSFP data of six subjects were acquired with a 12-point phase-cycling scheme and corresponding RF phase increments φ, uniformly distributed in the range (0, 360°): i.e., $\varphi_j = 180°/12 \cdot (2j-1)$, j = 1, 2, …12. Other bSSFP protocol parameters included an isotropic resolution of 1.3 x 1.3 x 1.3 mm$^3$, 128-144 partitions to cover the whole brain, a TR/TE of 4.8 ms/2.4 ms, a nominal flip angle $\alpha_{nom}$ of 15°, and 256 dummy preparation pulses before each phase-cycle acquisition. Using an in-plane generalized autocalibrating partial parallel acquisition (GRAPPA) acceleration factor of 2, the total scan time for phase-cycled bSSFP ranged from 10 min 12 s to 11 min 27 s depending on the number of partitions.

Diffusion-weighted images were acquired using a standard axial 2D multi-slice single-shot SE-EPI DTI sequence. Diffusion-sensitized scans were performed with a bipolar diffusion gradient scheme along 20 directions and a b-value of 1000 s/mm$^2$. Additionally, a non-diffusion weighted (b-value = 0) dataset was acquired with reverse phase encoding direction. Protocol parameters included a resolution of 1.4 x 1.4 x 3.0 mm$^3$, 36 slices for whole-brain coverage, a TR/TE of 4800 ms/83 ms, an in-plane GRAPPA acceleration factor of 2, 9 averages per direction and b-value, yielding a total acquisition time of 15 min 23 s.

The measurement protocol was complemented by a magnetization-prepared rapid gradient-echo (MPRAGE) sequence[50], which provided structural T$_1$-weighted images for segmentation purposes (isotropic resolution = 1.2 x 1.2 x 1.2 mm$^3$, TR/TE = 2200 ms/3.44 ms, $\alpha_{nom}$ = 8°, scan time = 3 min 7 s).

Data acquisition at 9.4 T

Whole-brain 3D sagittal phase-cycled bSSFP experiments were conducted in the same six subjects as scanned at 3 T. The employed 12-point bSSFP phase-cycling scheme was identical to the 3 T protocol. The acquisition was performed with a high isotropic resolution of 0.8 x 0.8

x 0.8 mm$^3$, 208-224 partitions to ensure whole-brain coverage, a TR/TE of 4 ms/2 ms, $\alpha_{nom}$ = 9°, a GRAPPA acceleration factor of 4 (in-plane: 2 / through-plane: 2), and 256 dummy preparation pulses before each phase-cycle acquisition, resulting in total scan times of 9 min 17 s to 9 min 58 s depending on the number of partitions.

Data processing pipeline

To achieve distortion correction of the 3 T diffusion data, the two non-diffusion weighted datasets with opposite phase-encoding directions were used to estimate the susceptibility-induced off-resonance field, which served as additional input for the subsequent eddy current correction of the 4D DWI data. Then, the diffusion tensor was calculated voxelwise based on weighted least-squares fitting to derive the eigenvalues $\lambda_1$, $\lambda_2$, and $\lambda_3$ as well as the principal diffusion eigenvector $V_1$ corresponding to the largest eigenvalue $\lambda_1$. From the eigenvalues, the scalar diffusion metrics MD, FA, AD = $\lambda_1$, and RD = $(\lambda_2 + \lambda_3)/2$ were obtained. The principal eigenvector $V_1$ in Cartesian coordinates (x, y, z) was transformed to spherical coordinates (r, $\Phi$, $\Theta$; with r = 1 per definition). The inclination $\Theta$ represents the angle between $V_1$ and $B_0$ while the azimuth $\Phi$ defines the angle between the positive x-axis and the projection of $V_1$ into the x-y plane. The six diffusion metrics MD, FA, AD, RD, $\Phi$, and $\Theta$ derived at 3 T served as target for NN training at both field strengths (cf. Fig. 1). To provide matched NN input data, the high-resolution 3 T and 9.4 T bSSFP phase-cycles as well as the 3 T anatomical MPRAGE were registered and downsampled to the lower-resolution 3 T DTI data using the distortion-corrected non-diffusion weighted dataset as reference. More details about the preprocessing of the bSSFP data are available online in the Supporting Information (subsection 1).

Brain extraction, i.e., skull-stripping, and subsequent brain tissue classification based on the MPRAGE contrast provided white matter (WM), gray matter (GM), and cerebrospinal fluid (CSF) probability masks. Voxels containing pure CSF were excluded from the voxelwise NN training to increase the weight of tissues and the brain masks were further thresholded based on the target diffusion metrics to remove erroneous values according to: MD / AD / RD $\in$ (0, 2.5·10$^{-3}$] mm$^2$/s, FA $\in$ [0.01, 0.99], $\Phi$ / $\Theta$ $\in$ (0, 90] °.

The measured WM bSSFP profiles were binned voxelwise for different $\Theta$ ranges after $B_0$-shift correction to visualize the degree of asymmetry depending on fiber tract orientation. For comparison, the bSSFP frequency response was simulated using the analytical steady-state equation[51] for the respective scan parameters as well as literature WM relaxation times[52] at 3 T and 9.4 T. The dependence of bSSFP profile asymmetries in WM on FA and $\Theta$ was analyzed by calculating the asymmetry index AI[32,33] for WM voxels, binned into different FA as well as $\Theta$ ranges: $AI(FA,\Theta) = (h_n - h_p)/(h_n + h_p)$, where $h_{n,p}$ refers to the signal peak at negative (n) and positive (p) frequency offsets in the magnitude of the $B_0$-corrected bSSFP profile relative to the banding. Binning included the whole-brain datasets of all subjects by retaining WM voxels classified with a probability higher than 90%. For each bin, the mean and standard deviation (SD) was computed.

NN architecture

In this proof-of-principle study, multilayer perceptrons, i.e., fully connected feedforward NNs were used as the basic architecture for voxelwise supervised learning as sketched in Figure 1. Since the MR signal evolution of one voxel may spread to its neighbors, e.g., due to residual registration errors, a 3x3 window in the axial plane was extracted around each voxel and used in a vectorized form as NN input. Real and imaginary parts of the 12-point complex bSSFP profile were stacked, yielding a total of 216 input features. The NN output consisted of twelve features represented by the six diffusion metrics MD, FA, AD, RD, $\Phi$, and $\Theta$ as well as their respective uncertainties $\sigma_i$ ($i$ = 1, 2, … 6) to make the NN network probabilistic[41] and allow posterior uncertainty estimation in future applications where likely no target data will be available. Target diffusion data for the supervised learning process were obtained from standard DTI measurements as described above. The uncertainties $\sigma_i$ were indirectly inferred during the training process based on the principle of maximum likelihood (see below) without requiring any labeled targets.

The depth of the multilayer perceptron is defined by the number of hidden layers (HLs) between input and output layers. The total NN size is determined by the number of neurons in each HL. Here, an identical number of neurons was assumed for all HLs. The neural input in the HLs was processed by a ReLU (rectified linear unit) activation function. For the neurons in the output layer, a linear activation function was used. To find an optimal NN architecture, hyperparameter optimization was performed based on 12-point bSSFP data acquired at 3 T by treating the number of HLs, the number of neurons per HL, and the batch size as variables as described in detail in the Supporting Information (subsection 2). It was assumed that 3 T and 9.4 T bSSFP profiles share sufficiently similar information content to justify the use of identical optimal NN models. The hyperparameter optimization returned an optimum for the sampled hyperparameter space of 4 HLs, 500 neurons per HL, and a batch size of 128, which was selected in the following as optimal NN architecture, yielding in total 866.012 learnable parameters for the model.

NN training

The training data contained ~590.000 voxels from four subjects; 20% of these voxels were used for validation purposes during training. Prior to training, the input data was standardized (i.e., mapped to mean = 0 / SD = 1) and the target data was normalized (i.e., scaled to the range [0, 1]). NN weights were adjusted using the Adam[53] optimization algorithm with a fixed learning rate of $10^{-4}$. An additional uncertainty term was added to the loss function to implicitly learn the posterior variance of the NN output parameters by applying the maximum likelihood principle and extending the standard mean squared error (MSE) to a negative log-likelihood loss function[41,42,54,55]

$$-\log p(\boldsymbol{\mu}^{tgt}; \boldsymbol{\mu}(\boldsymbol{x}), \boldsymbol{\sigma}(\boldsymbol{x})) = \frac{1}{2}\sum_{i=1}^{n}(\frac{\mu_i^{tgt}-\mu_i(\boldsymbol{x})}{\sigma_i(\boldsymbol{x})})^2 + \sum_{i=1}^{n}\log(\sigma_i(\boldsymbol{x})) + \frac{n}{2}\log(2\pi),$$

for a given input-target pair ($\boldsymbol{x}$, $\boldsymbol{\mu}^{tgt}$). This forces the NN to not only adjust the mean $\mu_i(\boldsymbol{x})$ but also the standard deviation $\sigma_i(\boldsymbol{x})$ of a Gaussian probability distribution during training. The means $\mu_i(\boldsymbol{x})$ and the SDs $\sigma_i(\boldsymbol{x})$ ($i = 1, 2, \ldots n = 6$) refer to the predictions of the six diffusion metrics and the corresponding uncertainties, respectively, for a given input bSSFP profile $\boldsymbol{x}$. Using the identified optimal NN architecture, separate networks were trained for 12-point bSSFP input data at 3 T and 9.4 T. For each field strength, the training was repeated 10 times with different random initializations of the weights.

NN testing on unseen data

Final NN outputs were derived for two unseen test subjects (subjects 1 and 2), not included in NN training, as the average over all 10 NN predictions. The NN-predicted maps were calculated for both, downsampled bSSFP data matched to standard DTI data to enable a direct validation of the NN quantification accuracy against the reference, and high-resolution bSSFP data to demonstrate the feasibility of deriving multiple quantitative diffusion metrics with whole-brain coverage at high isotropic resolution.

The predicted uncertainties were transformed voxelwise into a relative metric in units of percentage to allow for a direct comparison across different brain tissue structures and quantitative parameters: $\sigma_{i,rel} = (\sigma_i/\mu_i^{tgt}) \cdot 100$. The NN-predicted diffusion metrics at 3 T and 9.4 T were quantitatively validated in WM and GM masks containing pure WM and GM voxels (i.e., classified with 100% probability), respectively. The WM masks were further subdivided based on different FA ranges according to the reference DTI measurement, yielding the following five masks: $WM_1$, $WM_2$, $WM_3$, $WM_4$, and $WM_5$ with FA values in the range (0.0, 0.2], (0.2, 0.4], (0.4, 0.6], (0.6, 0.8], and (0.8, 1.0), respectively. In addition, smaller regions-of-interest (ROIs) were defined based on the acquired anatomical MPRAGE in WM (i.e., in a frontal region and in the corpus callosum (CC)) and in GM (i.e., in the putamen and the thalamus), cf. Supporting Information Figure S1. The quantitative accuracy of the predicted parameters was assessed for all masks/ROIs by the relative error in units of percentage defined as $\Delta_{rel} = [(<\mu_i> - <\mu_i^{tgt}>)/<\mu_i^{tgt}>] \cdot 100$ (< > refers to the mean in the respective masks). All reported means, SDs, and relative errors are pooled over the two test subjects.

**RESULTS**

WM bSSFP profile asymmetries

Balanced SSFP profile asymmetries in WM are highly dependent on the fiber tract orientation relative to $B_0$, i.e., on the inclination angle $\Theta$ (cf. Figs. 2A+D). At 9.4 T (Fig. 2D), enhanced asymmetries with stronger dependency on $\Theta$ can be seen in comparison to 3 T (Fig. 2A). A clear increase of the AI can be observed when incrementing $\Theta$ from parallel (0°) to orthogonal (90°) tract orientation (cf. Figs. 2B+E) with AI levels, which are almost twice as high at 9.4 T than at 3 T. In addition, the AI positively correlates with FA in WM, yielding overall higher AI levels for increasing FA values. On the other hand, the AI is almost unaffected by the azimuthal angle $\Phi$ (cf. Figs. 2C+F).

These findings can be visualized on a voxelwise basis by comparing AI maps at 3 T and 9.4 T (cf. Fig. 3A) with inclination $\Theta$ and FA maps (cf. Fig. 3B) for a WM mask overlaid onto the anatomical $T_1$-weighted MPRAGE. The AI maps at 3 T (Fig. 3A, left) and 9.4 T (Fig. 3A, right) reveal highest asymmetries in highly anisotropic WM structures (i.e., high FA values close to 1) and perpendicular fiber tract orientations relative to $B_0$ (i.e., $\Theta \approx 90°$), e.g., in the CC genu and splenium.

NN testing on unseen data

Training the optimized NN architecture with 10 random weight initializations for phase-cycled bSSFP input data at 3 T as well as 9.4 T and applying the 10 networks to test subject 1, resulted in a mean performance (i.e., sum of MSEs) of 0.200 ± 0.003 / 0.239 ± 0.004 for 3 T / 9.4 T, using the same brain masks as defined for NN training. The NN performance of 9.4 T data is at similar but slightly lower level as compared to 3 T data.

NN predictions of all six investigated diffusion metrics and their associated relative uncertainties are shown in Figure 4 for 3 T and 9.4 T bSSFP input data of test subject 1 not contained in NN training and downsampled to match the reference DTI data. A general structural smoothing and slight loss of contrast can be observed for all predicted quantitative parameters in comparison to the reference measurements, e.g., in the posterior internal capsule (IC) or the optic radiations (OR), cf. colored arrows in Figure 4. The results at both field strengths bear high resemblance. The relative uncertainty maps allow a fast visual assessment of the confidence level of the parameter prediction in the displayed representative axial slice, indicating that the NN prediction is generally reliable for MD (overall), AD (overall, but better in WM), RD (overall, apart from CC), FA (in WM, while GM exhibits higher uncertainties mainly due to division by low FA values), and inclination $\Theta$ (apart from a few structures, mainly in GM, which appear impaired, likely due to the discrete nature of the angle, which is difficult to be learned by the NN). The NN prediction of the azimuth $\Phi$ fails almost completely (apart from a few WM structures), presumably since this metric is not correlated with bSSFP profile asymmetries (cf. Figs. 2C+F) in contrast to $\Theta$ (cf. Figs. 2B+E), and will therefore not be validated quantitatively.

The NN predictions of MD are in high agreement with the reference measurement at both field strengths (cf. Fig. 4). MD is consistently slightly overestimated by the NN with mean relative errors below 9% in the assessed WM and GM masks (cf. Fig. 5A) and is barely dependent on FA as expected, i.e., similar results are yielded for $WM_{1-5}$. In contrast, a prominent dependence of the relative error on the WM anisotropy can be observed for FA (cf. Fig. 5B), AD (cf. Fig. 5C), and RD (cf. Fig. 5D). The NN model is able to correctly predict a linear dependence with increasing FA / AD values and decreasing RD values for increasingly anisotropic WM structures, but it underestimates the absolute value of the slope. In case of FA and AD, this results in an overestimation for structures with low FA (e.g., for $WM_1$) and an underestimation for structures with high FA (e.g., for $WM_5$) as can be seen in Figures 5B+C. In case of RD, an increasing overestimation for higher FA values is apparent (cf. Fig. 5D), also reflected by clearly higher relative uncertainties in the CC (cf. Fig. 4). The MD, FA, AD, and RD mean values of the whole-brain GM masks are overall in good agreement with the reference.

The ROI analysis in selected WM and GM structures yields generally good agreement between NN predictions and reference for MD, FA, AD, and RD with relative errors on the order of 10% or below (e.g., for MD) at both 3 T and 9.4 T; cf. Table 1. Increased errors occur in CC, which is characterized by a high anisotropy, leading to an underestimation of FA as well as AD and especially to a pronounced overestimation in RD, in accordance with the results presented in Figure 5.

The NN performance in predicting the inclination Θ is analyzed in Figure 6 for three different Θ ranges: (0, 30], (30, 60], and (60, 90]. For all three ranges, the mean relative errors in WM decrease with increasing FA while the NN prediction appears to fail for low FA values, especially for small Θ. The obtained low relative errors of the NN-predicted Θ for high FA are reflected by low uncertainties of Θ in highly anisotropic WM structures (cf. Fig. 4). The NN prediction of Θ in the GM masks yields unreliable results as apparent from the bar charts where similar mean values can be observed for all three Θ ranges.

Based on the validation of the NN-predicted diffusion metrics in Figures 5 and 6 as well as Table 1, the quantification accuracy can be considered similar for both investigated field strengths. The ability to deliver whole-brain quantitative maps of multiple diffusion metrics with high isotropic resolution of 1.3 x 1.3 x 1.3 $mm^3$ and 0.8 x 0.8 x 0.8 $mm^3$ is illustrated in Figures 7 and 8 for MD, FA, AD, and RD at 3 T and 9.4 T, respectively, using 12-point bSSFP phase-cycling schemes. The NN predictions exhibit a high-resolution structural information content, revealing fine-scale WM structures as indicated by the arrows.

# DISCUSSION

This proof-of-principle study indicates the high potential of NNs to predict high-resolution volumetric maps of multiple diffusion metrics jointly based on phase-cycled bSSFP input data at high to ultra-high field strengths (cf. Figs. 7 and 8). The trained NNs demonstrate ability to reliably estimate scalar diffusion measures (here, MD, FA, AD, RD) with isotropic resolutions of 1.3 x 1.3 x 1.3 mm$^3$ and 0.8 x 0.8 x 0.8 mm$^3$ at 3 T and 9.4 T, respectively, yielding whole-brain coverage in scan times of only about 10 min for 12-point phase-cycling schemes. The prediction of diffusion directionality information (here, the spherical angles $\Phi$ and $\Theta$) was impaired due to the discrete nature of these metrics and a weak correlation with the bSSFP asymmetry index in case of $\Phi$.

The NN prediction results are highly consistent across the two investigated field strengths (cf. Fig. 4). A possible explanation for the apparent structural smoothing in the NN-predicted parameter maps are interpolation effects since a general solution is parametrized to best approximate the target data during NN training. While the general structural information is preserved for the NN-predicted scalar diffusion metrics MD, FA, AD, and RD, fine-scale structural details, e.g., regions where fiber crossings reduce FA such as in the posterior internal capsule or the optic radiations (pointed to by colored arrows in Figs. 4, 7, and 8), are not captured by the NN predictions displayed in Figure 4, but they seem to reappear in the high-resolution maps (cf. Figs. 7 and 8). A certain extent of the smoothing visible in Figure 4 may thus be attributed to the downsampling of the bSSFP data to the low-resolution DTI reference data. The 3x3 nearest neighbor NN input was observed to slightly enhance the blurring in the parameter maps in comparison to a single-voxel input. However, it also yielded more stable parameter estimation results with reduced noise as it is less sensitive to image misregistrations and provides additional spatial information, which may be especially beneficial for brain regions affected by severe $B_0$ inhomogeneities.

Pronounced structural deficiencies in the NN predictions of the azimuthal angle $\Phi$ indicate that the orientation of the principal diffusion vector in the x-y plane is not directly contained in the phase-cycled bSSFP data, which hinders a better inference of this parameter. The finding that the AI does not reflect any information about $\Phi$ (cf. Figs. 2C+F) supports this interpretation. In contrast, a successful estimation of the inclination $\Theta$ in anisotropic WM structures (cf. Fig.

6) corroborates that enhanced bSSFP profile asymmetry in WM tracts perpendicular to $B_0$ (cf. Figs. 2B+E) provides useful information for the NN, reflected by lower relative uncertainties in these regions. Figure 5 reveals a high correlation between the strength of anisotropy in WM and the NN prediction performance of FA, AD, and RD. While the NN is able to correctly predict an increase of FA and AD as well as a decrease of RD with increasing anisotropy, it fails to properly estimate the absolute value of the slope of this nearly linear relationship, by underestimating it. Repeated measurements with slightly different head orientations and thus altered asymmetry levels in WM could be employed in future to improve the NN performance.

The use of complex data was beneficial as the phase contains information about the local frequency ($B_0$) and aided to increase the robustness of the NN predictions in the presence of $B_0$ field variations. However, residual banding-related artifacts in regions of high $B_0$ inhomogeneity such as near the sinuses were encountered in some cases. Therefore, we will aim at improving the handling of complex data in future work, e.g., by the development of complex-valued neural networks instead of splitting the complex-valued data into real and imaginary parts. This offers the benefit that the neural network model can incorporate the mathematical correlation between real and imaginary input channels directly during training, eliminating the need to learn phase information from scratch.

The employed voxelwise neural network fitting by means of a multilayer perceptron is inherently well suited for tissue quantification tasks as it learns the intrinsic dependence of the acquired signal evolution on multiple tissue characteristics. Under motion-free conditions, the MR signal series in one voxel can be considered independent of surrounding voxels. Nevertheless, to better learn the dependency on WM anisotropy, it appears useful to include spatial features into NN training. Convolutional neural networks (CNNs), which are well known for feature encoding and decoding of structural information, e.g., in the context of brain segmentation[56,57], could therefore be investigated in the future.

Since in practice, e.g., in clinical applications, no reference data will be available to evaluate the NN performance, probabilistic output layers were used in this work, providing voxelwise uncertainty estimation for each target parameter. Special care has to be taken in future studies when applying NNs to patient data since it is not *a priori* clear whether NNs trained in healthy volunteers are directly applicable to pathological conditions. Likely, retraining will be needed by including pathological data as input into NN training.

To the best of the author's knowledge, this is the first proposed method enabling simultaneous voxelwise mapping of diffusion parameters directly from phase-cycled bSSFP contrasts. The high-resolution NN-predicted maps in Figures 7 and 8 demonstrate the feasibility and reliability to jointly estimate multiple diffusion metrics with whole-brain coverage at 3 T and 9.4 T. The absence of diffusion gradients in the balanced SSFP acquisition has the inherent advantage to provide high motion robustness. In contrast, standard DTI sequences such as multi-shot SE-EPI are adversely affected by shot-to-shot phase errors caused by physiological motion, which necessitates the acquisition of navigators, which prolongs scan time. Due to the presence of strong $B_0$ field inhomogeneities inducing image distortions, SE-EPI-based diffusion-weighted scans are barely applicable at ultra-high fields. Furthermore, long echo trains in EPI acquisitions lead to $T_2^*$ decay along the phase encoding direction and consequently spatial blurring. Balanced SSFP imaging benefits from a comparably narrower point spread function, which allows resolving smaller voxels of submillimeter resolution at ultra-high field within scan times on the order of only 10 min for whole-brain coverage.

In addition, the presented NN-based approach does not require any complex mathematical modeling as compared to conventional diffusion tensor fitting, thus offering very fast multi-parametric quantification of diffusion metrics once the NN is trained. This adds considerable value to the proposed NN-driven approach, which is able to infer high-resolution distortion-free maps of various diffusion metrics with high structural information content from phase-cycled bSSFP input data within a processing time of only a few seconds.

It has to be noted that the proposed NN prediction of selected diffusion metrics based on the bSSFP frequency profile does not provide the same information content as standard DTI. As evidenced by the results, directionality information can likely only partially be derived since the estimation of the azimuthal angle Φ fails and therefore only a subset of diffusion measures with respect to DTI can be provided. However, the estimation of diffusion metrics could in future be synergistically combined with the quantification of relaxation parameters validated previously using analytical or NN models based on the same underlying phase-cycled bSSFP acquisition scheme[38–40], thus without any increase of scan time. This may offer the joint quantification of various clinically highly relevant tissue characteristics at high speed and high resolution.

## CONCLUSION

The presented NN-driven approach showed potential to jointly derive the diffusion metrics MD, FA, AD, RD, and Ө from phase-cycled bSSFP data. The trained NNs were able to provide high-resolution quantitative diffusion data of remarkable structural details with whole-brain coverage. The focus of future studies will be on eliminating the residual bias in the predicted diffusion metrics, e.g., by investigating complex-valued NNs and image-based CNNs.

## ACKNOWLEDGEMENTS

The financial support of the Max Planck Society and the German Research Foundation (DFG, Reinhart Koselleck Project, DFG SCHE 658/12) is gratefully acknowledged.

**TABLE**

|  |  | WM | | GM | |
|---|---|---|---|---|---|
|  |  | Frontal WM | Corpus Callosum | Putamen | Thalamus |
| **MD** (10⁻⁴) [mm²/s] | Reference | 7.7 ± 0.3 | 8.1 ± 0.8 | 7.1 ± 0.5 | 7.5 ± 0.8 |
|  | 3 T | 8.0 ± 0.1 | 8.2 ± 0.6 | 7.3 ± 0.2 | 7.7 ± 0.5 |
|  | 9.4 T | 7.9 ± 0.1 | 8.0 ± 0.5 | 7.6 ± 0.2 | 7.6 ± 0.5 |
|  | $\Delta_{rel}$ 3 T [%] | +4 | +1 | +3 | +3 |
|  | $\Delta_{rel}$ 9.4 T [%] | +3 | <1 | +7 | +1 |
| **FA** | Reference | 0.40 ± 0.05 | 0.80 ± 0.09 | 0.17 ± 0.07 | 0.31 ± 0.07 |
|  | 3 T | 0.47 ± 0.04 | 0.70 ± 0.08 | 0.19 ± 0.03 | 0.30 ± 0.06 |
|  | 9.4 T | 0.42 ± 0.04 | 0.63 ± 0.08 | 0.18 ± 0.04 | 0.27 ± 0.03 |
|  | $\Delta_{rel}$ 3 T [%] | +19 | -12 | +13 | -2 |
|  | $\Delta_{rel}$ 9.4 T [%] | +5 | -21 | +9 | -14 |
| **AD** (10⁻⁴) [mm²/s] | Reference | 11.2 ± 0.8 | 17.8 ± 1.3 | 8.3 ± 0.8 | 9.8 ± 0.9 |
|  | 3 T | 12.6 ± 0.6 | 16.1 ± 0.8 | 8.8 ± 0.5 | 10.3 ± 0.9 |
|  | 9.4 T | 11.9 ± 0.4 | 15.0 ± 1.5 | 9.0 ± 0.5 | 9.8 ± 0.6 |
|  | $\Delta_{rel}$ 3 T [%] | +13 | -10 | +6 | +5 |
|  | $\Delta_{rel}$ 9.4 T [%] | +6 | -16 | +9 | <1 |
| **RD** (10⁻⁴) [mm²/s] | Reference | 5.9 ± 0.4 | 3.2 ± 1.2 | 6.5 ± 0.6 | 6.4 ± 0.9 |
|  | 3 T | 5.7 ± 0.2 | 4.2 ± 0.9 | 6.6 ± 0.2 | 6.4 ± 0.5 |
|  | 9.4 T | 6.0 ± 0.3 | 4.5 ± 0.8 | 6.9 ± 0.2 | 6.5 ± 0.5 |
|  | $\Delta_{rel}$ 3 T [%] | -4 | +31 | +1 | +1 |
|  | $\Delta_{rel}$ 9.4 T [%] | <1 | +42 | +6 | +3 |

**Table 1.** ROI analysis of MD, FA, AD, and RD quantification accuracy pooled over both test subjects, which were not included in NN training, for relevant structures in WM (frontal WM, corpus callosum) and GM (putamen, thalamus) as defined in Supporting Information Figure S1. The NN predictions at 3 T and 9.4 T based on 12-point phase-cycled bSSFP input data are compared to the reference diffusion metrics derived with standard SE-EPI DTI. Positive and negative relative errors indicate an overestimation and underestimation with respect to the reference, respectively.

**FIGURES**

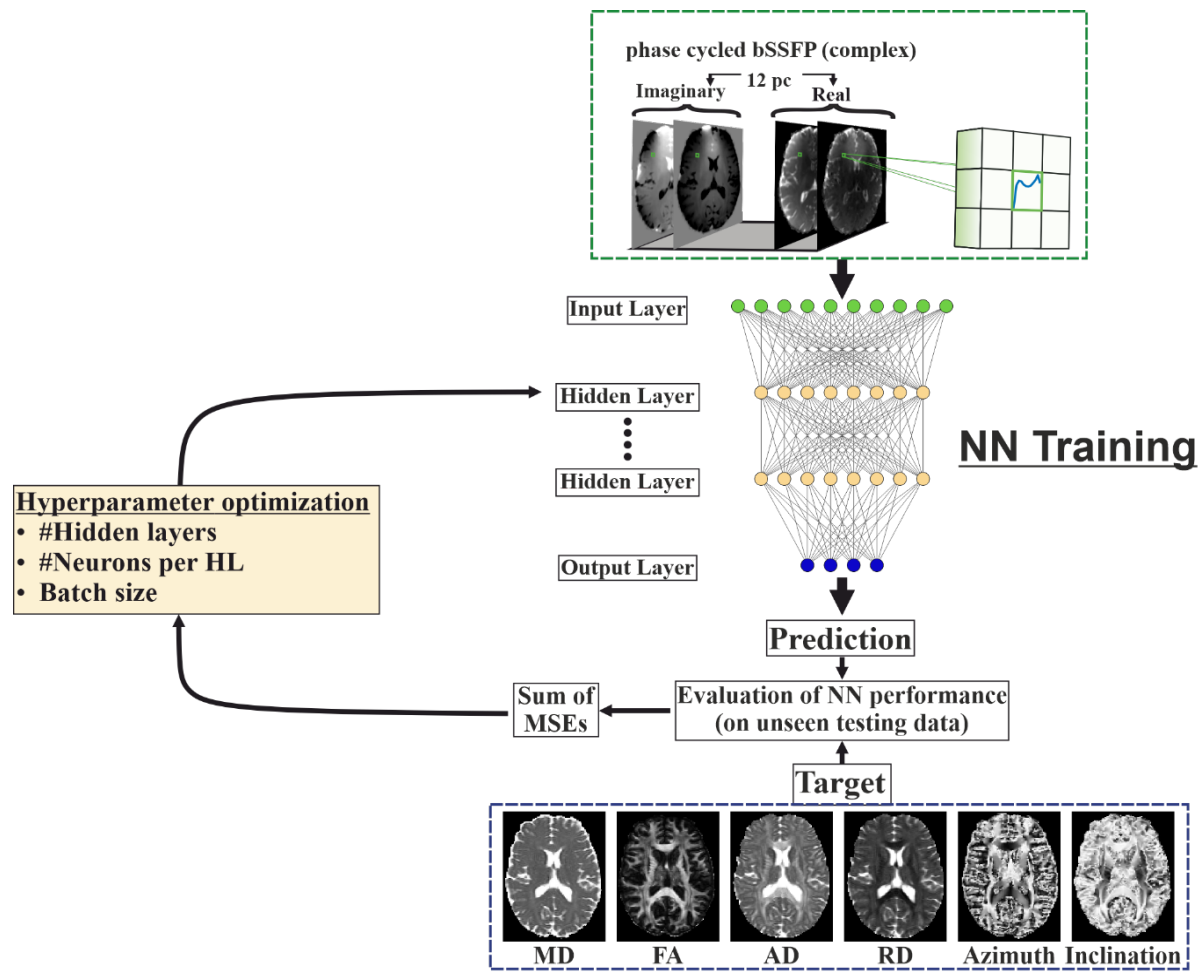

**Figure 1.** Neural network training and hyperparameter optimization scheme for the proposed multi-parameter mapping pipeline. Real and imaginary parts of the 12-point bSSFP phase-cycling data (green box) are fed voxelwise into a fully connected feedforward NN. For each voxel, a 3x3 window of nearest neighbors in the axial plane is extracted leading to 216 input features. After NN training, the predictions for unseen test data are compared to the target data obtained with standard SE-EPI DTI measurements (blue box) to evaluate the NN performance. The sum of all MSEs between predicted and targeted values was used as optimization metric for hyperparameter optimization (for more details about the hyperparameter optimization, cf. Supporting Information, subsection 2).

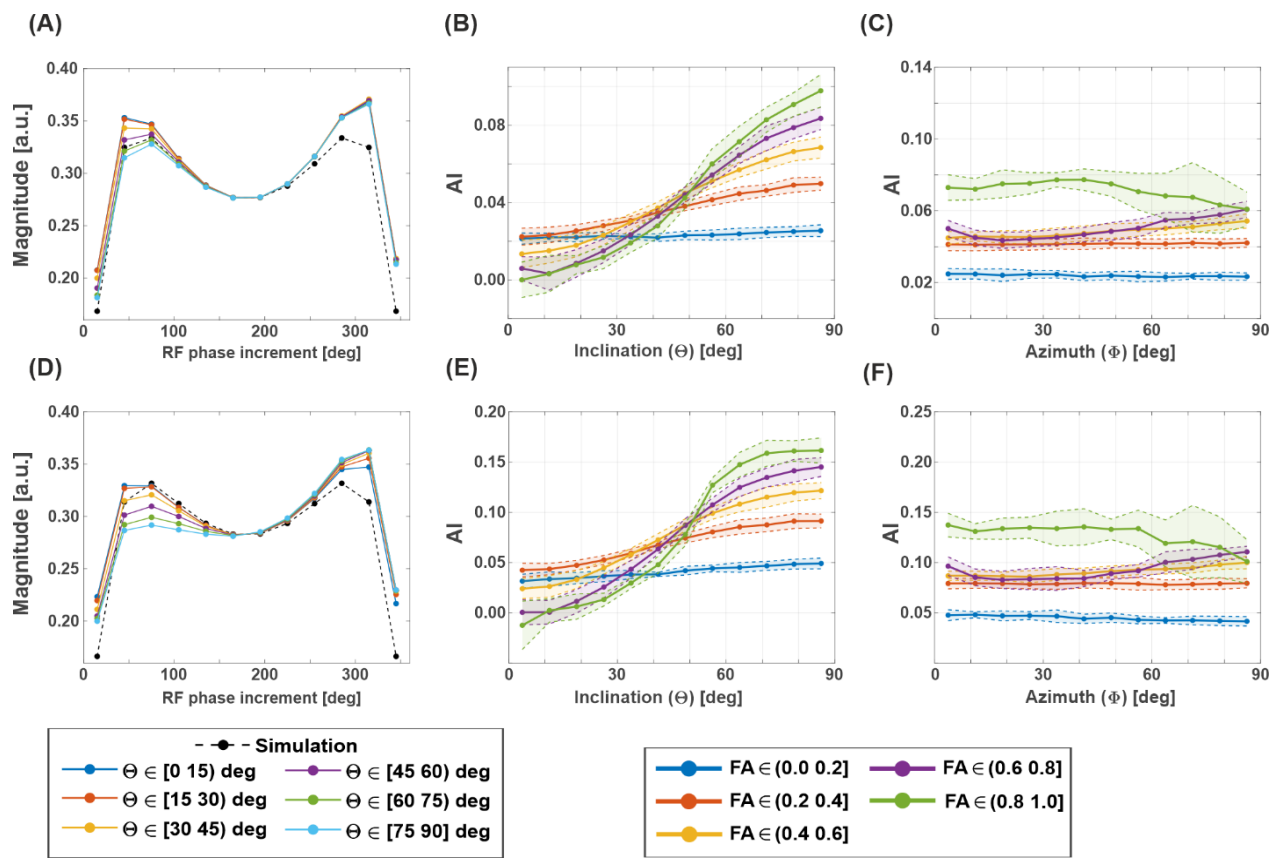

**Figure 2.** Balanced SSFP profile asymmetries and corresponding AI levels assessed at 3 T (**A-C**) and 9.4 T (**D-F**) in WM. (**A+D**) The magnitude of the measured 12-point bSSFP profiles in WM is binned voxelwise across all measured subjects for six different ranges of the inclination angle (Θ) as indicated in the legend and displayed for the average in each bin. Simulated analytical bSSFP profiles for literature WM relaxation times at 3 and 9.4 T are shown for comparison (black dashed curves). (**B+E**) and (**C+F**) The bSSFP asymmetry index (AI) in WM is plotted versus the inclination angle (Θ) (**B+E**) and the azimuthal angle (Φ) (**C+F**), respectively. The AI values are binned across all measured subjects and averaged for five different ranges of the fractional anisotropy (FA) as indicated in the legend with a bin size of 7.5° for both Θ and Φ. The shaded regions indicate the extent of ± SD (standard deviation of the mean for each bin).

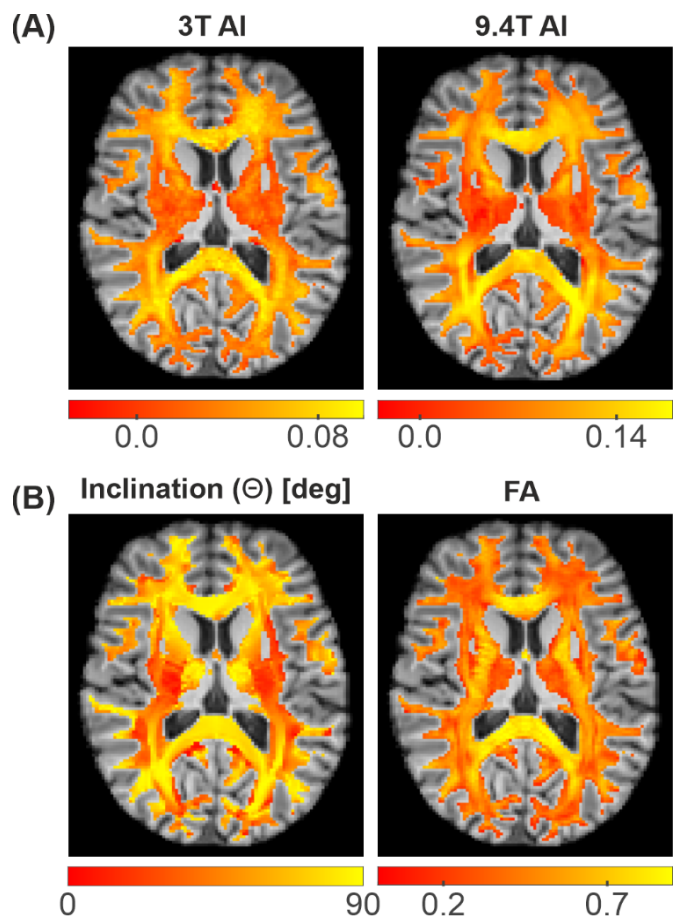

**Figure 3.** (**A**) Asymmetry index (AI) maps obtained at 3 T (left) and 9.4 T (right) for a representative axial slice in the same subject, registered to and overlaid on the structural $T_1$-weighted MPRAGE acquired at 3 T. Please note the different scaling of AI for 3 T and 9.4 T since the AI levels observed at 9.4 T are almost twice as high as the corresponding AI values at 3 T. (**B**) Maps of the inclination angle (Θ) (left) and the fractional anisotropy (FA) (right) obtained with standard SE-EPI DTI measurements at 3 T overlaid on the $T_1$-weighted MPRAGE.

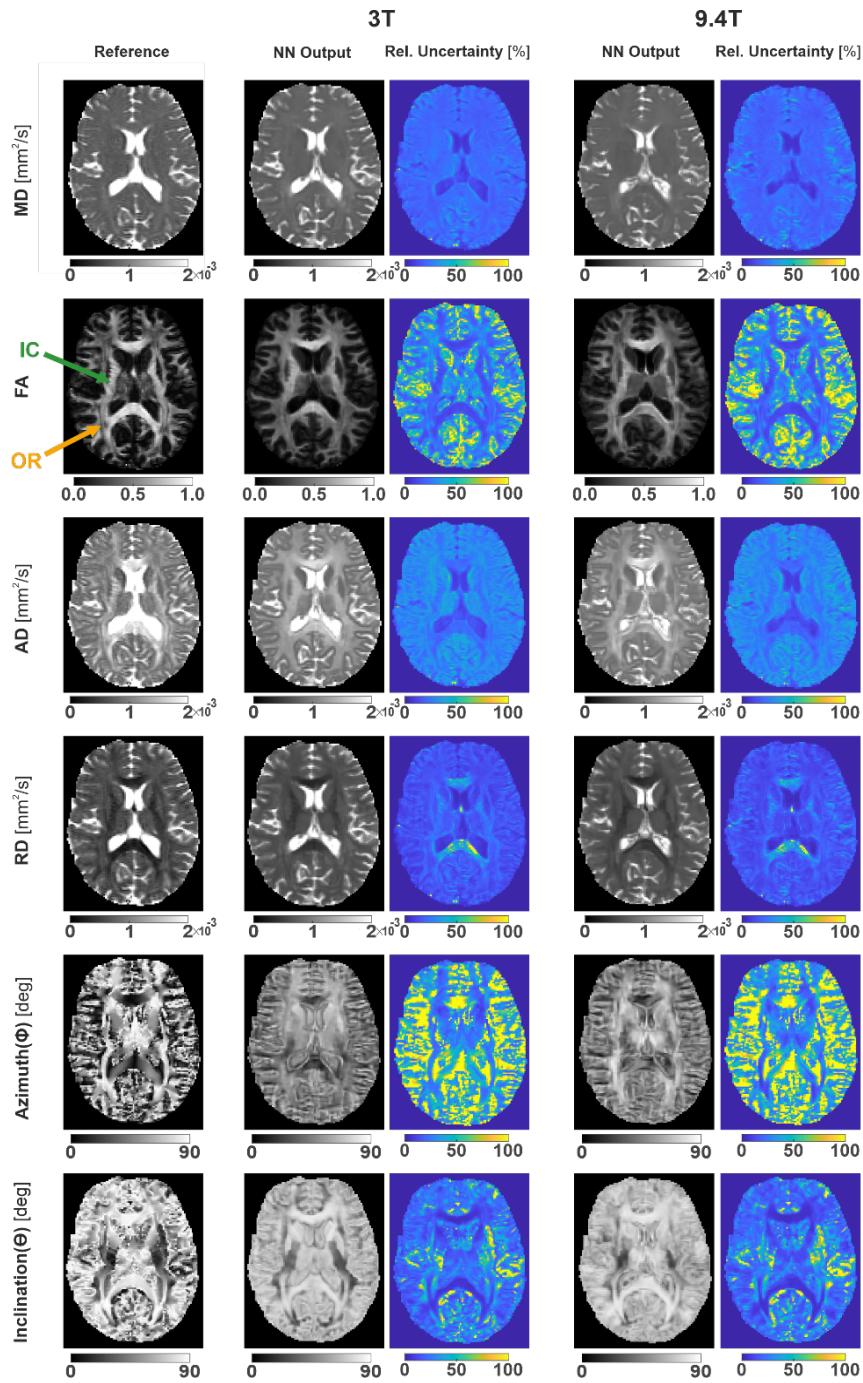

**Figure 4.** NN predictions are depicted for 12-point phase-cycled bSSFP data of test subject 1, not included in NN training, acquired at field strengths of 3 T (middle) and 9.4 T (right) in comparison to reference data obtained with standard SE-EPI-based DTI fitting (left). A representative axial slice is shown. The green and orange arrows drawn on the reference FA map indicate the location of the posterior internal capsule (IC) and the optic radiations (OR), respectively. Both 3 T and 9.4 T bSSFP data were co-registered (including downsampling) to the reference data. Relative uncertainty maps allow comparing the confidence level of the NN parameter prediction in different brain tissue structures and across quantitative parameters.

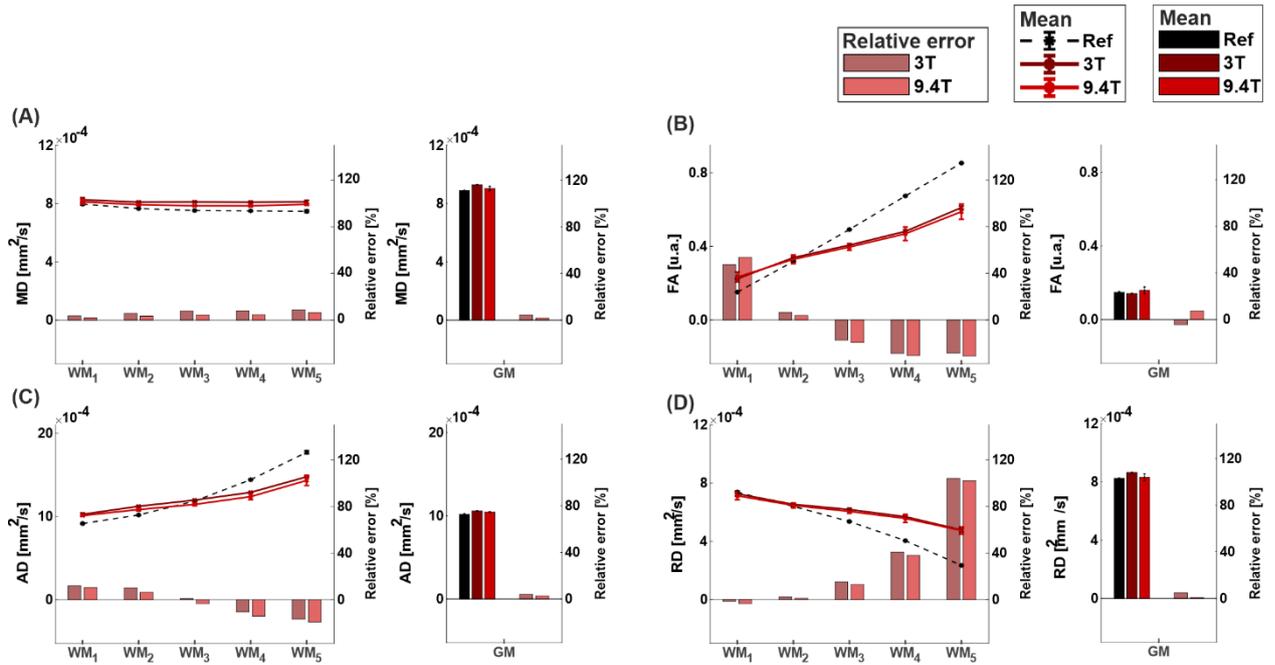

**Figure 5.** WM and GM mask analysis of the NN quantification accuracy in MD, FA, AD, and RD at 3 T and 9.4 T for 12-point bSSFP input data. The masks $WM_1$, $WM_2$, $WM_3$, $WM_4$, and $WM_5$ refer to WM voxels with FA values in the range (0.0, 0.2], (0.2, 0.4], (0.4, 0.6], (0.6, 0.8], and (0.8, 1.0), respectively. Gray matter is assessed by a whole-brain mask (GM). The relative error in percentage (right axis) between NN predictions and reference is depicted by light red-colored bar charts. Negative and positive values refer to a mean underestimation and overestimation, respectively, relative to the reference in the respective brain tissue mask. The mean values of the quantitative parameters (left axis) for the assessed masks are displayed in black and dark red colors for the reference and NN, respectively, as dots connected by straight lines for better visibility across the WM mask and as bar charts for GM. For both, relative error and mean, darker red refers to 3 T and lighter red refers to 9.4 T.
28

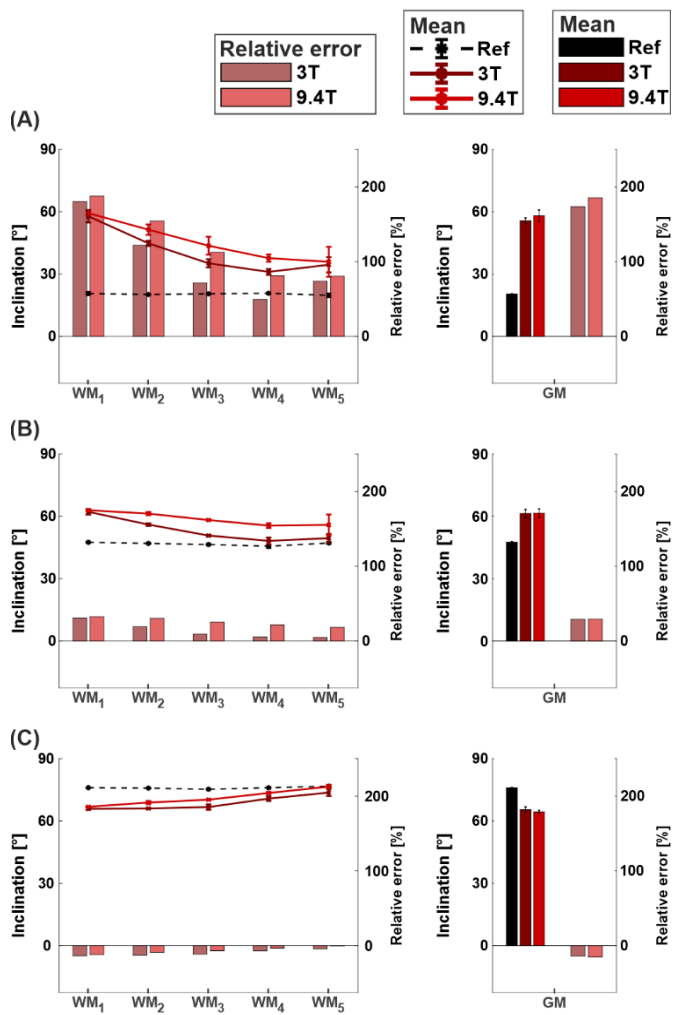

**Figure 6.** WM and GM mask analysis of the NN quantification accuracy in the inclination Θ at 3 T and 9.4 T for 12-point bSSFP input data. The WM and GM masks are defined as in Figure 6. Additionally, the masks are subdivided into three different Θ ranges according to the reference DTI measurement: (0, 30] (**A**), (30, 60] (**B**), and (60, 90] (**C**). The relative error in percentage (right axis) between NN predictions and reference is depicted by light red-colored bar charts. Negative and positive values refer to a mean underestimation and overestimation, respectively, relative to the reference in the respective brain tissue mask. The mean Θ values (left axis) for the assessed masks are displayed in black and dark red colors for the reference and NN, respectively, as dots connected by straight lines for better visibility across the WM mask and as bar charts for GM. For both, relative error and mean, darker red refers to 3 T and lighter red refers to 9.4 T.

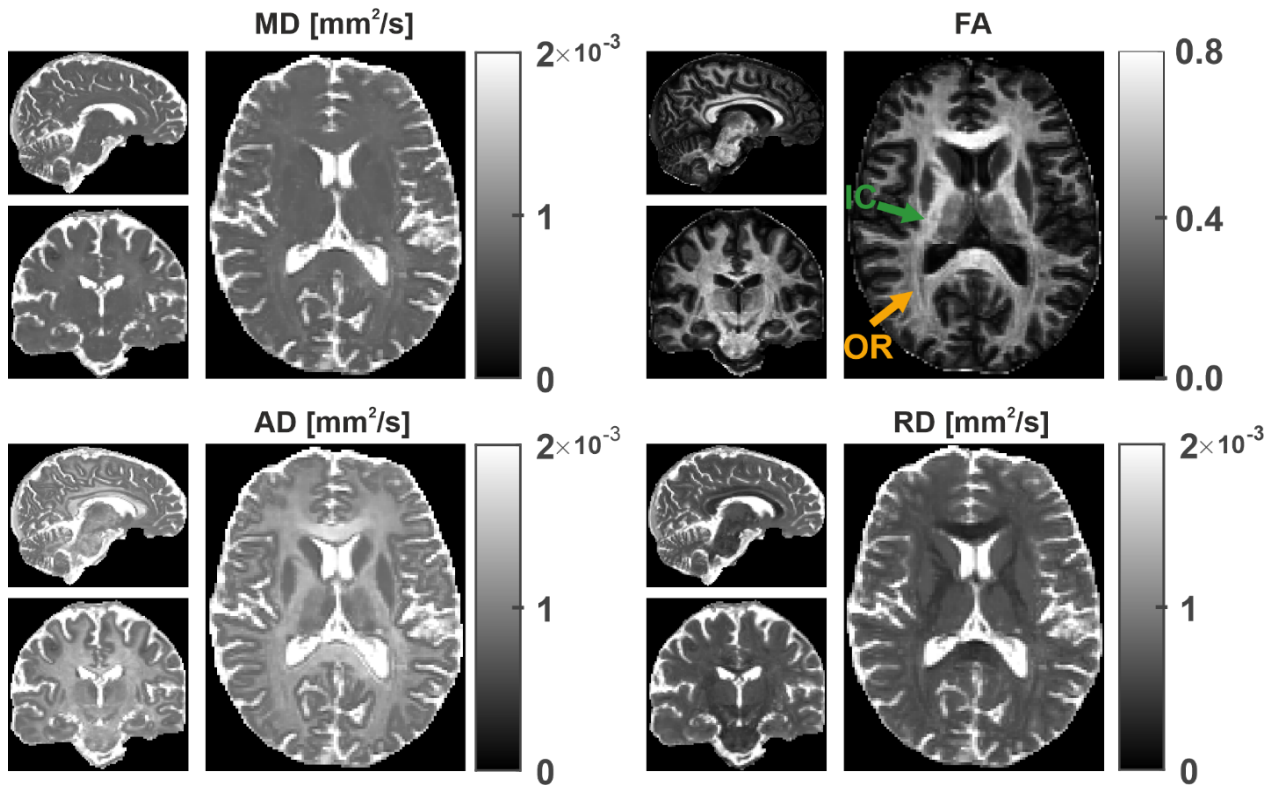

**Figure 7.** High-resolution whole-brain NN predictions of MD, FA, AD, and RD are depicted for 12-point phase-cycled bSSFP data of test subject 1, not included in NN training, acquired at a field strength of 3 T with an isotropic resolution of 1.3 x 1.3 x 1.3 mm$^3$. Representative sagittal, coronal, and axial slices are displayed for each parameter. Please note that the FA maps are windowed differently here as compared to Figure 4 to improve the visual appearance and enhance the high-resolution structural information content. The green and orange arrows in the axial slice of the FA prediction indicate the location of the posterior internal capsule (IC) and the optic radiations (OR), respectively.

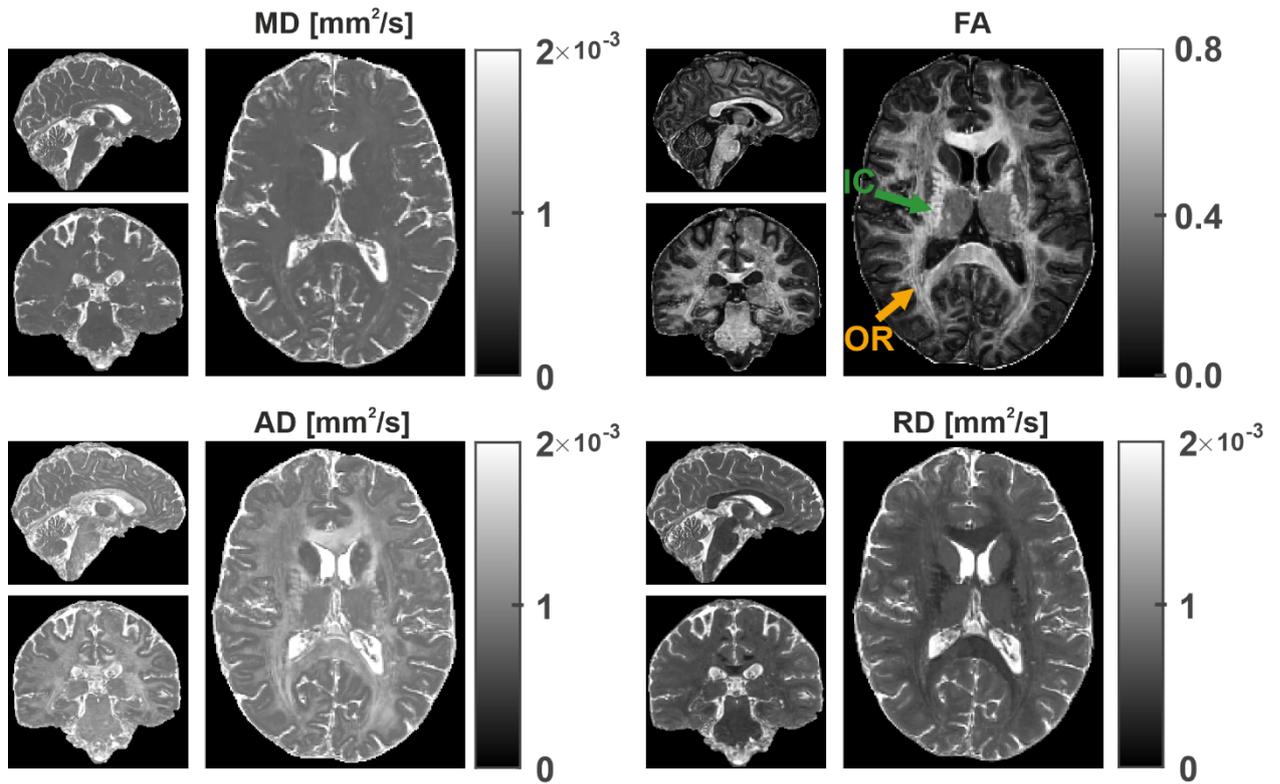

**Figure 8.** Ultra-high field high-resolution whole-brain NN predictions of MD, FA, AD, and RD obtained at 9.4 T with an isotropic resolution of 0.8 x 0.8 x 0.8 mm$^3$ are depicted for the same subject (test subject 1) and similar slice positionings as in Figure 7. Please note that the FA maps are windowed differently here as compared to Figure 4 to improve the visual appearance and enhance the high-resolution structural information content. The green and orange arrows in the axial slice of the FA prediction indicate the location of the posterior internal capsule (IC) and the optic radiations (OR), respectively.

**SUPPORTING INFORMATION**

*1. Preprocessing of bSSFP data*

The preprocessing of the 3 T and 9.4 T bSSFP data was identical. First, the phase-cycled bSSFP data were corrected for receiver-related phase offsets and then, ringing artifacts in the reconstructed images were removed based on local subvoxel shifts[1]. Subsequently, the twelve 3D bSSFP volumes were registered onto each other to eliminate potential misalignments due to bulk head motion during the scan. The coregistration process relied on the magnitude data by taking the magnitude of the 6th bSSFP phase-cycle ($\varphi = 165°$) as reference. The obtained image transformation matrices were applied to the respective phase data.

*2. NN hyperparameter optimization*

To perform hyperparameter adjustment with efficient sampling of the hyperparameter space, the optimization function *gp_minimize* from the *scikit-optimize* library[2] was applied, which performs Bayesian optimization using Gaussian processes. The NN hyperparameter space ($Y$), here the number of HLs (in the range [1, 5]), the number of neurons per HL (in the range [10, 500]), and the batch size (in the range [32, 512)], power of 2), were tuned to find the optimal model, which best approximates the input-to-target mapping. For a given hyperparameter setting ($y \in Y$), a NN was trained using a maximal number of 1000 epochs with a validation patience of 100 and the NN status at minimal validation loss was considered optimal to prevent overfitting. The plurality of possible hyperparameter combinations is computationally expensive to evaluate and can be considered as a black-box derivative-free function $g(y)$[3–7]. Global minimization of the objective function $g(y)$ was performed using the total sum of MSEs between predictions and target values as optimization metric.

Efficient sampling of the hyperparameter space and estimation of a surrogate model for $g(y)$ was anticipated within 100 iterations, using Bayesian optimization. The first hyperparameter set was provided by the user, followed by 10 random initializations. These first 11 observations were used to determine a prior probability distribution of $g(y)$ based on Gaussian processes. Subsequently, with each new hyperparameter setting ($y \in Y$), Gaussian processes describe a Bayesian posterior probability distribution, referred to as surrogate model, which is updated after each new observation. The computation time and results for the optimization process mentioned in this work refer to an NVIDIA Quadro RTX 5000 GPU with 16 GB of graphics processor memory.

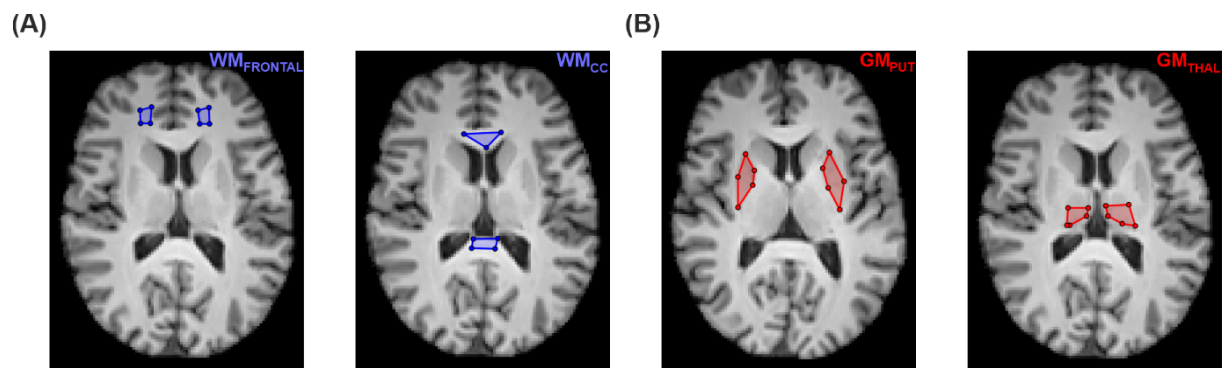

**Supporting Information Figure S1.** Definition of regions-of-interest (ROIs) used to validate the accuracy of NN-predicted MD, FA, AD, and RD values in representative axial slices, which were selected for two testing subjects not included into NN training. ROIs were drawn in two white matter structures (A), i.e., in a frontal region ($WM_{FRONTAL}$) and in corpus callosum ($WM_{CC}$), as well as in two gray matter structures (B), i.e., in putamen ($GM_{PUT}$) and thalamus ($GM_{THAL}$), as shown exemplarily for subject 1.